\documentclass[aps,preprint,showpacs,floatfix,preprintnumbers,amsmath,amssymb]{revtex4}
\usepackage{dcolumn}
\usepackage{bm}
\usepackage{graphicx}

\begin{document}
\title{The convergence method to calculate particles fluxes in x rays spectrometry techniques. Application in nuclear compounds}

\author{C. Figueroa, N. Nieva, H. Brizuela, S. P. Heluani}
      \affiliation{Laboratorio de F\'{i}sica del S\'{o}lido,  Dpto. de F\'{i}sica, Facultad de Ciencias Exactas y Tecnolog\'{i}a,
Universidad Nacional de Tucum\'{a}n, Argentina.}

\date{\today}

\begin{abstract}
A method to calculate particle fluxes applicable in most of the spectroscopy techniques is described.  Flux intensities of backscattered or absorbed electrons and emitted photons are calculated using a method of convergence to solve the Invariant Embedding equations that are used to describe the particle trajectories  inside a solid sample. Our results are found to be helpful to carry out a procedure for quantitative  characterization using instrument such as Electron Probe Microanalysis or other probes. Examples of application to calculate the composition of ternary alloys are given and are compared with the same calculations using another procedure.
\end{abstract}


\maketitle

\bigskip

\section{Introduction}
 In most of the spectroscopy techniques, the signals produced by the different probes must be recorded and interpreted considering the  different kinds of interactions between particles during its trajectory through an analyzed sample. This procedure involves hard calculations of fluxes of particles like electrons and/or photons, neutrons, protons etc. These fluxes must be obtained solving a transport problem where some particles produce other particles: as an example, electrons transport could produce electrons, photons or both of them as a consequence of their interactions with matter. The transport problem, in the spectroscopy area, is usually solved using Monte Carlo Calculation  \cite{Volker:97, Reed:93, Garth:ANS04,vickerman:09} or numerical solutions of the the time-independent Boltzmann transport equation. Although considerably advances were made in the subject, most of the methods used in the practice have an empirical or semi empirical origin \cite{Penelope, Case:67, Werner:PRB97, Bote:JESRP09, Novak:PRB09, Werner:PRB11}.\\
 The quantitative determination of the unknown chemical composition of a material is an essential goal for its characterization. Among the spectroscopy techniques, the most used by researchers in materials science for compositional determinations is the Electron Probe Microanalysis (EPMA). EPMA is based on the detection of the characteristic X-rays emitted from a solid sample when a beam of electrons impinges on  its surface  with energies that typically vary between 2 and 50 keV. This instrument combines the capabilities of both the scanning electron microscope  and an X-ray fluorescence spectrometer. The electrons of the beam are scattered as they penetrate within the sample producing backscattered electrons, secondary electrons, absorbed electrons and characteristic and continuum X-rays spectra. Quantitative microanalysis in EPMA is performed by comparing the characteristic X-ray intensity from one element in the sample with that from a reference standard containing a known amount of the same element, by means of the so called the $K$ ratio,
  \begin{equation}
     K=\frac{I}{I_{p}}
                    \end{equation}
 Where  $I$ and $I_{p}$ are the intensities of the characteristic X-rays  emitted from the analyzed sample  and that from a standard.   \cite{Castaing:thesis, Reed:93, SIA:SIA04}.
  The $K$ ratio  could be transformed into the mass concentration of the element of interest taking into account some correction factors  due to processes undergone by the characteristic X-ray and by the electrons of the beam inside the sample. These processes involve scattering dynamics related to the sizes of the different atomic nuclei present in the sample and the standard, absorption, fluorescence. The respective correction factors arising from each of these processes are called \textit{Z},\textit{A} and  \textit{F} in the literature \cite{Reed:93}. Then, the atomic fractions of the elements are obtained using the following expression:
\begin{equation}
K=\frac{I}{I_{p}}= ZAF \frac{C}{C_{p}}
\end{equation}
Where $C$ and $C_{p}$ are the atomic fractions of the element of interest in the sample and in the standard.

   The most common procedures to estimate the corrections are the so called \textit{Matrix Correction} which are based on the calculation of the X-ray production as a function of the sample depth,  $\varphi (\rho z)$ to estimate both the absorption and the atomic number corrections \cite{Packwood:XRS82,Pouchou85, Reed:93}. Most of the analytical expressions  for the $\varphi (\rho z)$ functions used in commercial software packages are obtained from ad hoc models. Also the expressions for the $\varphi (\rho z)$ function and for the backscattering electron coefficients are consider as unrelated\cite{Heluani:XRS05}.
On the other side, in the literature there are theoretical models based on Boltzmann's transport equation to describe the phenomenon and to  interpret the signals in EPMA \cite{Oswald:JESRP93, Werner:PRB05}. Although, this procedure has the disadvantage that there are solutions only for the one dimension problem, and little progress has been made on the 2-D or 3-D cases. Therefore this has a reduced practical application. Monte Carlo calculations is still the most worthwhile method of approaching theoretical problems in EPMA \cite{Jablonski:PRB89,Penelope, SIA:SIA04, Aleksander2005}. However, depending on the accuracy or resolution required, running times can be computationally expensive.

 In the last two decades a few authors demonstrated the efficiency of the Invariant Embedding (IE) method to describe the scattering of electrons inside a solid sample\cite{Heluani:NIM02, Heluani:Xray03, Glazov:PRB03, Glazov04, Glazov2007, Figueroa:JAP06, Pazic09, Figueroa:JMS10}. This method is useful to estimate the flux of particles through the interfaces in a material medium as a function of experimental parameters and the detected signals. It is, the IE method avoids calculations of fluxes inside the analyzed samples, thus gaining calculation advantages, and yields a set of equations for the detected fluxes as a function of the size of the analyzed sample. This set can be solved as an initial-value problem. Particularly, the backscattering problem was successfully solved by different authors using IE method \cite{Heluani:NIMB00, Glazov:PRB03}. However, a procedure based on IE, useful for the experimentalist to perform chemical characterization, is still lacking.

In this article we focus on the calculation of the chemical compositions of a materials by using the experimental $K$ ratios and theoretical X-ray intensities obtained with IE.  Although the IE method is a powerful tool in order to simplify the transport equations, in 3-D transport problems the addition of degrees of freedom of the particles generates more complex mathematic. The Method of Convergence (MC)  proposed here allows us to deduce closed approximated solutions for the 3-D IE equations for the detected X-rays  intensities. These intensities can be calculated with no need of previous knowledge of the $\varphi(\rho z)$ function because the absorption corrections are already included in the equations.
Here we propose two solutions for the IE equations with different levels of approximation. One called the \textit{simple expression} and the other, more rigorous, called the \textit{complete expression}. Comparison with other model proves that both solutions are a suitable tool for the calculation of intensities and concentrations.

\section{Theoretical model}
The states ladder model and MC were are described in previous reports\cite{Figueroa:JAP06}. Basically,  after impinging with energy E0, electrons can suffer two kinds of interactions: elastic or inelastic. Elastic collisions only cause deviations while inelastic ones only produce energy losses. Based on the assumption that it is allowed only one kind of inelastic collision with a constant energy loss of value $\Delta$E, a discrete set of possible values is generated: E1, E2, E3 and E4, for a five level system. For energies E $<$ E4, electrons are deactivated.
Characteristic radiation can be generated exclusively by the inelastic collisions. Contributions I01, I02, I03, I04 to the total intensity are made by electrons backscattered with E1, E2, E3 and E4, while contribution I4 corresponds to the absorbed electrons. All these intensities can be calculated with the IE method. This method starts with the deduction of the integro-differential equation for each contribution to the intensity, whose derivation variable is the thickness $\tau$ of  the sample.
 In the case of $I_{\alpha_{0},\beta_{1}}$, the intensity contribution for the electrons that impinge with E0 and $\alpha_{0}$ and are backscattered with E1 and $\beta_{1}$, the following equation is obtained:

\begin{eqnarray}
\frac{d I_{\hat{\alpha_{0}},\hat{\beta_{1}}}(\tau)}{d\tau}=\frac{g_{0}R_{\hat{\alpha_{1}},\hat{\beta_{1}}}(\tau)s_{0}}{\hat{\alpha_{0}}}  \nonumber\\ +\frac{g_{0}R_{\hat{\alpha_{0}},\hat{\beta_{0}}}(\tau)s_{0}}{\hat{\beta_{1}}}-I_{\hat{\alpha_{0}},\hat{\beta_{1}}}(\frac{2(s_{0}+\sigma_{0})} {\hat{\alpha_{0}}}+\mu)+\nonumber\\
\int_{\varphi}\frac{d\sigma_{\hat{\alpha_{0}},\hat{\varphi}}I_{\hat{\varphi_{0}},\hat{\beta_{1}}}(\tau)}{\hat{\alpha_{0}}}+
\int_{\hat{\epsilon}}\frac{d\sigma_{\hat{\epsilon_{1}},\hat{\beta_{0}}}I_{\hat{\alpha_{0}},\hat{\epsilon_{1}}}(\tau)}{\hat{\epsilon_{1}}}+\nonumber\\
\int_{\epsilon}\frac{d\sigma_{\hat{\epsilon_{1}},\hat{\beta_{0}}}I_{\hat{\alpha_{0}},\hat{\epsilon_{1}}}(\tau)}{\hat{\epsilon_{1}}}+
\int_{\epsilon}\int_{\eta}\frac{d\sigma_{\hat{\epsilon_{0}},\hat{\eta_{0}}}R_{\hat{\alpha_{0}},\hat{\epsilon_{0}}}(\tau)I_{\hat{\eta_{0}},\hat{\beta_{1}}}(\tau)}
{\hat{\epsilon_{0}}}+\nonumber\\
 \int_{\gamma}\int_{\delta}\frac{d\sigma_{1,\gamma,\delta}R_{\hat{\delta_{1}},\hat{\beta_{1}}}(\tau)I_{\hat{\alpha_{0}},\hat{\gamma_{1}}}(\tau)}{\hat{\gamma_{1}}}.
\label{eqnI01}
\end{eqnarray}

 Were the simplified notation $\hat{\alpha_{j}}=cos (\alpha_{j})$ and $\hat{\beta_{j}}=cos (\beta_{j})$ is used for the  incident and emerging director cosines of the polar angles of the electrons, measured with respect to the normal of the the sample surface. The index $j=0,1,..$ labels the energy states values,$\sigma_{j}$ and $s_{j}$ are the elastic and inelastic cross sections,respectively; $\varphi$, $\epsilon$ and $\delta$  are the scattering angles due to  single interactions inside the sample, and $g_{j}$ is the electrons efficiency of  X-ray production per collision.

Using the MC it is possible to obtain two approximate solutions for the last equation: the complete form, i.e., the formal solution, and the simple form, used initially as an auxiliary expression to solve the equations of the lower levels. These approximate solutions tend to an exact solution for $\tau \rightarrow \infty$.The simple form is:

\begin{equation}
I_{\hat{\alpha_{0}} \hat{\beta_{1}}}(\infty)=\frac{g_{0}\hat{\alpha_{0}}\hat{\beta_{1}}s_{0}}{(\lambda+1)(s_{1}\hat{\alpha_{0}}+s_{0}\hat{\beta_{1}})+\mu\hat{\alpha_{0}}
\hat{\beta_{1}}}\frac{(1+\lambda)\Gamma_{1}}{\lambda J_{0}}
\end{equation}

 Where $J_{0}=\lambda +2-\alpha \Gamma_{1}$; $\Gamma_{1}=\lambda=2-2\sqrt{\lambda+1}$ and $\lambda=\sigma_{0}/s_{0}$.

In order to study the applicability range of the solutions, chemical composition calculations made with the simple and the complete expressions are presented.
The factor of generation $gj$ it calculated starting from Bethe's expression for $L_{k}$ lines, and an energy level E$_{j}$, for the ionized element in the compound:
\begin{equation}
 g_{kj}=12.672 10^{-8} \frac{\rho_{k}N_{A}}
 {A_{k}E_{j}I_{k}}Ln(\frac{E_{j}}
{1.6 10^{-16}l_{k}})\sqrt{\frac{2E_{j}}{m}}
\end{equation}

Where $_{k}$  is the partial density , A$_{k}$is the atomic number and m is the mass of the electron.
Parameters such as the partial densities, ionization energies or attenuation coefficients are obtained from tables. The cross sections are calculated from classic expressions as functions of the partial densities. In this model, it is considered that the beam electrons interact with atoms in the compound, whose cross sections is the pondered average of the cross sections of each element.
The procedure is as follows: 1) The summation of the measured \textit{K} ratios is normalized. 2) This normalized values are considered as the initial atomic concentrations. 3) With these new initial atomic fractions, the \textit{K} ratios are recalculated with the simple and the complete MC expressions, and the results are compared with the measured values. 4) In successive calculations, the atomic fractions of both expressions are adjusted in order to fit the theoretical and experimental \textit{K} ratios. The final values for the concentrations are those that yield the best fit.  The calculation was performed for the line $L_{\alpha}$  of Zr and$K_{\alpha}$ of Fe.

\section{Characteristic intensities. Simple expressions.}

We proceed to calculate I01, I02, I03, I04 and I4 in order to obtain $I=\sum_{i} I0i$

The simple expressions for the Intensities are:
\begin{eqnarray}
I01(\infty)=\int_{0}^{1}\frac{4g_{0}a_{0}s_{0}\hat{\beta_{1}}(1+\lambda)T_{1}d\hat{\beta_{1}}}
{\{(\lambda+1)
(s_{1}\hat{\alpha_{0}}+s_{0}\hat{\beta_{1}})+\mu \hat{\alpha_{0}}\hat{\beta_{1}}\}\lambda J_{0}}
 \end{eqnarray}
 \begin{eqnarray}
I02(\infty)=\int_{0}^{1}\frac{8g_{2}a_{0}s_{0}(1+\lambda)T_{1}W_{0}\hat{\beta_{2}}d\hat{\beta_{2}}}
{[(s_{2}\hat{\alpha_{0}}+s_{0}\hat{\beta_{2}})(1+\lambda)+\mu \hat{\alpha_{0}}\hat{\beta_{2}}]\lambda J_{0}^{3}}
 \end{eqnarray}
 \begin{eqnarray}
I03(\infty)= \nonumber \\
\int_{0}^{1}\frac{16g_{3}a_{0}s_{0}(1+\lambda)T_{1}\hat{\beta_{3}}d\hat{\beta_{3}}
 \{W_{0}R_{0}+b_{0}c_{0}Z_{0}(\lambda+2)\}}{(s_{3}\hat{\alpha_{0}}+s_{0}\hat{\beta_{3}})(1+\lambda)+\mu \hat{\alpha_{0}}\hat{\beta_{3}}
 \lambda J_{0}^{4}}
 \end{eqnarray}
 \begin{eqnarray}
  I04(\infty)=
  \int_{0}^{1}\frac{32 g_{4}a_{0}s_{0}T_{1}(\lambda+1)\hat{\beta_{4}}}{\{\lambda J_{0}^{6}[s_{4}\hat{\alpha_{0}}+s_{0}\hat{\beta_{4}}]\}}\nonumber \\
  \times \{b_{0}c_{0}J_{0}Z_{0}(d_{0}V_{0}+\lambda+2)+
W_{0}(a_{0}b_{0}T_{1}+J_{0})+ \nonumber \\ a_{0}b_{0}T_{1}\{W_{0}R_{0}+c_{0}Z_{0}[b_{0}(\lambda+2)+d_{0}V_{0}+W_{0}]\}\}d\hat{\beta}_{4}
 \end{eqnarray}

 where

 \begin{eqnarray}
 Z_{0}=\frac{a_{0}(b_{0}-2)T_{1}+2\lambda +4}{\lambda+2+2\sqrt{\lambda+1}-a_{0}T_{1}}\\
 W_{0}=(2+\lambda)(b_{0}+1)-a_{0}T_{1}
 \end{eqnarray}

\section{Assessment of the results: Zr-based alloys }
In order to compare the calculated K ratios and concentrations between MC and PAP \cite{Pouchou84,Pouchou87,Pouchou91} results, we used experimental data obtained with an analytical microprobe of electrons CAMECA SX50 with 20 kV of accelerating potential from samples of ternary metallic alloys of the systems Zr-Sn-Fe and Zr-Sn-Nb. For the calibration, high purity standards of each of the analyzed elements were used. The proposed method was applied to a set of fourteen measurements on these alloys.
The nominal atomic composition of the Zr-Sn-Fe sample is: Zr 62.5 - Sn 25.0 - Fe 12.5. It was thermally treated to 800C for 2000 hours. Three phases are detected in this sample: two binary intermetallic compounds (ZrFe$_{2}$ and Zr$_{5}$Sn$_{3}$) and a ternary compound Zr$_{6}$Sn$_{2}$Fe. At first, the atomic composition was estimated using the PAP procedure. On the other hand, the theoretical results of section II and III were used to estimate the theoretical \textit{K} ratios and the atomic composition with the complete and the simple expressions for the characteristic intensities.  The results are shown in tables I and II.
They show comparisons between PAP concentrations and results obtained with the simple and the complete 3D-IE expressions, for the Zr$_{5}$ Sn$_{3}$ and Zr$_{6}$Sn$_{2}$Fe phases. For each: element of interest, measured \textit{K} ratio (\textit{k}), concentrations from the PAP method (at(PAP)), concentrations from 3D-IE simple equation (at(3Ds)),concentrations from 3D-IE complete equation (at(3Dc)), \textit{K} ratios from 3D-IE simple equation (\textit{k}s), \textit{K} ratios from 3D-IE complete equation (\textit{k}c). Each table contains different sets of values, corresponding to different measuring points in the same phase. The first rows of each set of values indicate the PAP, 3Ds and 3Dc estimations of the oxygen concentration, as this elements is always present at the surface of these alloys.
\begin{table}
  \caption{\label{tab:table1}Theoretical calculations (IE) for the K ratios of the elements (E) in the Zr$_{5}$Sn$_{3}$ phase of the Zr-Sn-Fe alloy. The 3D-IE results obtained with the complete (c) and the simple (s) expression are compared  with PAP calculation.}
  \begin{ruledtabular}
\begin{tabular}{lccccccc}
\hline
  && 3.16$\%$ O & 2.6$\%$ O & 15.1$\%$ O& &\\
 & \textit{K} & $\%$at(PAP) & $\%$at(3Ds) & $\%$ at (3Dc)&\textit{K$_{s}$(t)}& \textit{K$_{c}$(t)}\\
Zr & 0.4498& 61.3297& 53.1501&60.9755&\textbf{0.4460}& 0.4791\\
Sn&0.3303&38.2826&46.4461 &38.4317&\textbf{0.3268}&0.3418\\
Fe&0.0020&0.378&0.4038& 0.5928&\textbf{0.0020}&0.0020
\end{tabular}
  \begin{tabular}{lccccccc}
   && 6.28$\%$ O & 1.1$\%$ O & 13.1$\%$ O& &\\
Zr & 0.4527 & 61.7237& 53.93&61.4206&\textbf{0.4573}&0.4883 \\
Sn&0.3336&37.9375&45.6761 &38.0250&\textbf{0.3243}&0.3486\\
Fe&0.0018&0.3387&0.3557& 0.5544&\textbf{0.0018}&0.0019\\
\end{tabular}
\end{ruledtabular}
\end{table}

\begin{table}
    \caption{\label{tab:table2}Theoretical calculations (3D-IE) for the K ratios of the elements (E) and its atomic fractions in the  Zr$_{6}$Sn$_{2}$Fe phase of the Zr-Sn-Fe alloy. The 3D-IE results are compared  with PAP calculation.}
  \vskip 0.5cm
  \begin{ruledtabular}
  \begin{tabular}{lccccccc}
\hline
  && 1.14$\%$ O & 7.0$\%$ O & 7.0$\%$ O& &\\
 & \textit{K} & $\%$at(PAP) & $\%$at(3Ds) & $\%$ at (3Dc)&\textit{K$_{s}$(t)}& \textit{K$_{c}$(t)}\\
Zr & 0.5716 & 67.36&61.2063&65.294& \textbf{0.5561}&0.5982 \\
Sn&0.2120&22.7&27.7656 &21.6365&\textbf{0.2090}&0.2205\\
Fe&0.0604&9.925&11.028& 13.071&\textbf{0.0605}&0.0625
\end{tabular}
\begin{tabular}{lccccccc}
&& 1.47$\%$ O & 5.5$\%$ O& 7.0$\%$ O&\\
Zr & 0.5763& 67.43& 62.1693&65.3326&\textbf{ 0.5692 }&0.5990\\
Sn &0.2115 &22.5& 27.2487&21.5054& \textbf{0.2065} &0.2193\\
Fe &0.0616& 10.05& 10.5820&13.1720& \textbf{0.0585} &0.0631
\end{tabular}
\begin{tabular}{lccccccc}
&&1.18$\%$O& 7.4$\%$O&15.4$\%$ &\\
Zr& 0.5102&65.0721& 58.1445&63.4064& \textbf{0.4985} &0.5228\\
Sn &0.2680& 30.1344& 36.6928 &29.8788&\textbf{0.2621}& 0.2797\\
Fe &0.0272&34.8033 &5.16274&6.7148& \textbf{0.0268}&0.0285
\end{tabular}
\begin{tabular}{lccccccc}
&&2.13$\%$O&4.0$\%$O&7.5$\%$&\\
Zr&0.5625&66.5781& 60.6771 &64.5946&\textbf{0.5561}&0.5870\\
Sn& 0.2230&23.7969 &29.1667 &22.7027&\textbf{0.2217}& 0.2304\\
Fe &0.0585&9.6352& 10.1562& 12.7027&\textbf{0.0562}& 0.0603
\end{tabular}
\begin{tabular}{lccccccc}
&&0.21$\%$O&2.3$\%$ O&9.0$\%$&\\
Zr& 0.5629&66.4305& 60.0338&64.4540 &\textbf{0.5783}& 0.5784\\
Sn& 0.2292&24.3489& 29.9064& 23.3171&\textbf{0.2340}& 0.2341\\
Fe &0.0562& 9.2206& 10.0596& 12.2288&\textbf{0.0572}9& 0.0573\\
\end{tabular}
\end{ruledtabular}
\end{table}

On the other hand, the nominal atomic compositions of the Zr-Sn-Nb samples are: Zr 70.0 - Sn 5.0 - Nb 25.0 for the first one, and Zr 65.0 - Sn 10.0 - Nb 25.0 for the other one. They were thermally treated to 950C for 2900 hours. Table III  shows comparisons of the theoretical results in this article and experimental data, in the same way as Tables I and II.
\begin{table}
\caption{\label{tab:table3}Theoretical calculations for  $K_{\alpha}$  ratios of elements (E) in the  ZrSnNb alloy. .}
  \vskip 0.5cm
  \begin{ruledtabular}
  \begin{tabular}{lccccccc}
\hline
 & \textit{K} & $\%$at(PAP) & $\%$at(3Ds) & $\%$ at (3Dc)&\textit{K$_{s}$(t)}& \textit{K$_{c}$(t)}\\
Zr& 0.6071&64.50& 58.92&59.02& \textbf{0.6129}& 0.5946\\
Sn &0.0220& 2.59 &1.74& 2.44&\textbf{0.0217}& 0.0228\\
Nb& 0.3253&32.91& 39.34& 38.54&\textbf{0.3275}& 0.3155
\end{tabular}
\begin{tabular}{lccccccc}
Zr& 0.616& 66.03 &63.62&61.32& \textbf{0.6337}& 0.6034\\
Sn &0.0678&7.81& 5.72& 6.74&\textbf{0.0696}& 0.0671\\
Nb& 0.2552&26.17 &31.66 &31.94&\textbf{0.2556}& 0.2529\\
\end{tabular}
\begin{tabular}{lccccccc}
Zr& 0.5957&64.88& 59.02&59.02&\textbf{ 0.6126} &0.5934\\
Sn &0.0271&3.26& 2.16& 2.84&\textbf{0.0269}& 0.0269\\
Nb &0.3076&31.86& 38.82& 38.14&\textbf{0.3224}&0.3115\\
\end{tabular}
\begin{tabular}{lccccccc}
& & &\textbf{Sample 2}& & &\\ \\

Zr& 0.7291&78.39& 74.77& 74.77&\textbf{0.7464}&0.7309\\
Sn &0.0380& 4.46& 3.21 &4.09&\textbf{0.0325}& 0.0382\\
Nb &0.1677&17.14& 22.02&21.14& \textbf{0.1746}&0.1629\\
\end{tabular}
\begin{tabular}{lccccccc}
Zr &0.6931&70.82& 67.01& 67.01&\textbf{0.6831}& 0.6661\\
Sn &0.0294&3.51 &2.99& 2.99&\textbf{0.0292}& 0.0278\\
Nb &0.2489&25.67& 30.00 &30.0&\textbf{0.2487}& 0.2390
\end{tabular}
\begin{tabular}{lccccccc}
Zr &0.7071& 73.64 &68.78 &68.10&\textbf{0.6993}&0.6762\\
Sn &0.0269 &3.10& 2.20& 2.90&\textbf{0.0268}& 0.0269\\
Nb& 0.2344 &23.26& 29.02 &29.0&\textbf{0.2351}& 0.2304
\end{tabular}
\begin{tabular}{lccccccc}
Zr& 0.6849&72.92& 68.10& 68.10&\textbf{0.6932}&0.6762\\
Sn &0.0277&3.25& 2.25& 2.90&\textbf{0.0274}&0.0269\\
Nb &0.2352&23.255& 29.65& 29.0&\textbf{0.2405}&0.2304\\
\end{tabular}
\end{ruledtabular}
\end{table}

\section{Discussion}
Calculations performed with the simple and the complete expressions obtained with MC are compared with results of the PAP method. The fitting is acceptable keeping in mind the conditions of calculation and the fact that the intensity expressions were deduced from approximate expressions for the transport coefficients of the electrons. Thereby, the intensities are a second stage of approximation.
In the Zr-Sn-Fe results, the presence of oxygen changes the composition of the phases in a significant way. Oxygen diminishes the \textit{K} ratios of the other elements, in comparison with non-oxidized samples.
In relation to the phase Zr$_{5}$Sn$_{3}$,  it is necessary to remark that the results yielded by the complete MC expressions fit the PAP concentrations better than the simple ones. However, the complete expressions overestimate the oxygen concentrations at the surface.
In relation to the MC calculated \textit{K} ratios, the discrepancy is less than 5$\%$,  with respect to the PAP values. In general, for this compound, the simple MC expressions underestimate the Zr and overestimate the Sn, in comparison with PAP and complete MC expressions.
 It is necessary to emphasize that the equations are valid if the ratio $\psi= \mu/s_{0 }$ can be considered as negligible with respect to unity. In the calculations,$\Psi$ values are: 0.05$\%$ for Fe, 0.18$\%$ for Zr and 0.14$\%$ for Sn. It means that the MC expressions are at the limit of it applicability for the present selection of parameters.
In relation to the oxygen, the simple MC expressions overestimate its concentration in comparison with PAP.
The results for  Zr$_{6}$Sn$_{2}$Fe show the same behavior that those of the previous phase in the sense that there is a better fit between concentrations calculated with the complete MC expressions and PAP in comparison with the simple ones. These expressions yield results for the \textit{K} ratios that differ from the experimental values in an amount that ranges from 2 to 5 $\%$.
 In general, the theoretical \textit{K} ratios fit to experimental values within less than 3 $\%$, with an overestimation in the results of the simple MC expressions. The $\psi$ ratios for this phase are: 0.05 for Fe, 0.17 for Zr and 0.14 for Sn, so the MC expressions are at its limit of application. With regard to oxygen, PAP  concentrations are about 1$\%$ while MC values range from 7 to 15$\%$. In this phase the discrepancies in the concentrations of Zr, Sn  and Fe between PAP and MC range from 2 to 5$\%$.
In the case of Zr-Sn-Nb, the fit between K ratios is better, but the discrepancy between PAP and MC concentrations is higher. Both MC expressions underestimate the Zr and overestimate the Nb in comparison with PAP results. There is no oxygen at the surface of this compound, and the $\Psi$ ratios are: 0.12$\%$ for Nb, 0.16$\%$ for Zr and 0.17$\%$ for Sn.
    It is important to remark that the inelastic-elastic correlation determines the ionization rate and the $\phi(\rho z)$ function. Thereby, the intensities depend on such correlation, which is related to the $\lambda$ factor, this is to say, to the ratio between the inelastic and elastic cross sections.
Our calculations were performed basically to verify the internal coherence of the 3D model and the MC expressions. This means that parameters such as cross-sections, attenuation coefficients, ionization energies, etc are fixed values taken from tables, and the only adjustable parameter is the Emin value,i.e.,the minimum energy lost per inelastic collision, whose magnitude was selected to fit the backscattering and absorption coefficients of the electrons. Bearing in mind this fact, the presented results for concentrations and K ratios are quite satisfactory and indicate that the 3D model and the MC expressions are appropriate for the theoretical representation of the interaction phenomena in EPMA, and thereby, for the characterization of compounds by means of this or similar  techniques.
\begin{acknowledgments}
This work  was partially supported by CIUNT under grant
26/E439 and by ANPCyT under grants-PICT2448.
\end{acknowledgments}

\appendix
\section{Determination of the function $I_{\alpha_{0},\beta_{1}}(\infty)$. Method of Convergence.}
Considering in the integral-differential equation for $I_{\alpha_{0},\beta_{1}}(\infty)$, Eq.(\ref{eqnI01}) the six first terms equals to a constant I1:
\begin{eqnarray}
I1=&&\frac{g_{0}R00_{\alpha\beta}s_{0}}{\hat{\beta}_{1}}
\nonumber \\
&&+\frac{g_{0}R11_{\alpha\beta}s_{0}}{\hat{\alpha}_{0}}
+\int_{\varphi}\frac{d\sigma_{0\alpha\beta} I01_{\varphi\beta}}{\hat{\alpha}_{0}}\nonumber \\
&&+\int_{\xi}\frac{d\sigma_{1\xi\beta} I01_{\alpha\xi}}{\hat{\xi}_{1}}+\int_{\gamma}\int_{\delta}\frac{d\sigma_{1\gamma\delta}I01_{\alpha\gamma}R11_{\delta\beta}}{\hat{\gamma}_{1}}+\nonumber \\
&&\int_{\zeta}\int_{\eta}\frac{d\sigma_{0\xi\eta}R00_{\alpha\zeta}I01_{\eta\beta}}{\hat{\zeta}_{0}}
\end{eqnarray}
Eq.(\ref{eqnI01}) is now written as
\begin{equation}
\frac{dI_{\alpha_{0},\beta_{1}}(\tau))}{d\tau}=I1
-I_{\alpha_{0},\beta_{1}}(\tau)(\frac{s_{0}+\sigma_{0}}{\hat{\alpha}_{0}}+\frac{s_{1}+\sigma_{1}}{\hat{\beta}_{1}}+\mu)
\end{equation}
The solution for $\tau\rightarrow\infty\ $  is:
\begin{equation}
 I_{\alpha_{0},\beta_{1}}(\infty)=\frac{I1\hat{\alpha}_{0}\hat{\beta}_{1}}{\hat{\alpha}_{0}(s_{1}+
\sigma_{1})+\hat{\beta}_{1}(s_{0}+\sigma_{0})+\mu\hat{\alpha_{0}}\hat{\beta_{1}}}
\end{equation}

Let us rename the term  $I1\times\hat{\alpha_{0}}=$B1, were B1=B1($\hat{\alpha}_{0},\hat{\beta}_{1}$):
\begin{equation} I_{\alpha_{0},\beta_{1}}(\infty)=\frac{B1\hat{\beta}_{1}}{\hat{\alpha}_{0}(s_{1}+
\sigma_{1})+\hat{\beta}_{1}(s_{0}+\sigma_{0})+\mu\hat{\alpha}_{0}\hat{\beta}_{1}}
\end{equation}

Equation [A1] contains four integrals:

\begin{eqnarray}
\int_{\varphi}\frac{d\sigma_{0\alpha\beta} I_{\varphi_{0},\beta_{1}}(\infty)}{\hat{\alpha}_{0}}\\
\int_{\xi}\frac{d\sigma_{1\xi\beta}I_{\alpha_{0},\xi_{1}}(\infty)}{\hat{\xi}_{1}}\\
\int_{\gamma}\int_{\delta}\frac{d\sigma_{1\gamma\delta}I_{\alpha_{0},\gamma_{1}}(\infty)R_{\delta_{1}\beta_{1}}}{\hat{\gamma}_{1}}\\
\int_{\zeta}\int_{\eta}\frac{d\sigma_{0\xi\eta}R_{\alpha_{0}\zeta_{0}}I_{\eta_{0}\beta_{0}}}{\hat{\zeta}_{0}}
\end{eqnarray}

These integrals are solved by replacing $I_{\alpha_{0},\beta_{1}}(\infty)$ by the expression [A4] obtained with the MC.
The functions of Elastic Backscattering R00 are replaced by its normalized simple expression, deduced with the MC[reference].
Replacing at [A1],solving, multiplying for $\alpha_{0}$  and replacing $\rightarrow \Psi_{0}=\mu/s_{0}$ and $\Psi_{1}=\mu/s_{1}$ we obtain.

\begin{eqnarray}
 B1=\frac{2 g_{0} a_{0} s_{0} T_{1}}{\lambda}+\frac{\lambda s_{0} B1 \hat{\beta}_{1}}{2 s_{1} [\lambda+1+\Psi_{1}\hat{\beta}_{1}]} \nonumber \\
 \times\ln(\frac{s_{1}}{\hat{\beta}_{1}s_{0}}+1+\frac{\Psi_{0}}{\lambda+1})
+\frac{a_{0} s_{1} B1 T_{1}\hat{\alpha}_{0}\hat{\beta}_{1}}{s_{0}[\lambda+1+\Psi_{0} \hat{\alpha}_{0}]} \nonumber\\
 \times \ln(\frac{s_{0}}{\hat{\alpha}_{0}s_{1}}+1+\frac{\Psi_{1}}{\lambda+1})
 \ln(1+\frac{1}{\hat{\beta}_{1}})
+\frac{a_{0}s_{0}B1 T_{1}\hat{\alpha}_{0} \hat{\beta}_{1}}{s_{1} [\lambda+1+\Psi_{1}\hat{\beta}_{1}]} \nonumber \\
\times \ln(\frac{s_{1}}{\hat{\beta}_{1}s_{0}}+1+\frac{\Psi_{0}}{\lambda+1})\ln(1+\frac{1}{\hat{\alpha}_{0}})\nonumber\\+\frac{\lambda s_{1} B1 \hat{\alpha}_{0}}{2 s_{0}  [\lambda+1+\Psi_{0}\hat{\alpha}_{0}]}ln(\frac{s_{0}}{\hat{\alpha}_{0}s_{1}}+1+\frac{\Psi_{1}}{\lambda+1})\nonumber
\end{eqnarray}

Substituting this results in  [A4]  we obtain the following  expression for I01:
\begin{eqnarray}
 I01_{\alpha\beta}(\infty)^{c}=\frac{\hat{\beta}_{1}}{\hat{\alpha}_{0}(s_{1}+\sigma_{1})+\hat{\beta}_{1}(s_{0}+\sigma_{0})+\mu\hat{\alpha}_{0}\hat{\beta}_{1}}\nonumber\\
\{\frac{2 g_{0} a_{0} s_{0} T_{1}}{\lambda}+\lambda s_{0}\kappa_{0}g_{0}\{\frac{s_{0}\lambda \hat{\beta}_{1}}{2 s_{1} [\lambda+1+\Psi_{1}\hat{\beta}_{1}]} \ln(\frac{s_{1}}{\hat{\beta}_{1}s_{0}}\nonumber\\
+1+ \frac{\Psi_{0}}{\lambda+1})+\frac{a_{0} s_{1} T_{1}\hat{\alpha}_{0}\hat{\beta}_{1}}{s_{0}[\lambda+1+\Psi_{0} \hat{\alpha}_{0}]} \ln(\frac{s_{0}}{\hat{\alpha}_{0}s_{1}}+1+\frac{\Psi_{1}}{\lambda+1}) \nonumber \\ \times \ln(1+\frac{1}{\hat{\beta}_{1}})
 +\frac{a_{0} s_{0} T_{1}\hat{\alpha}_{0} \hat{\beta}_{1}}{s_{1} [\lambda+1+\Psi_{1}\hat{\beta}_{1}]} ln(\frac{s_{1}}{\hat{\beta}_{1}s_{0}}+1+
 \frac{\Psi_{0}}{\lambda+1})\nonumber \\
  \times \ln(1+\frac{1}{\hat{\alpha}_{0}})+
 \frac{\lambda s_{1} \hat{\alpha}_{0}}{2 s_{0} [\lambda+1+\Psi_{0}\hat{\alpha}_{0}]}ln(\frac{s_{0}}
 {\hat{\alpha}_{0}s_{1}}+1+\frac{\Psi_{1}}{\lambda+1})\}\nonumber \\
\end{eqnarray}
Where the super index "c" it refer to "complete".

Let us now substitute  $B1 =\kappa_{0}g_{0}\lambda s_{0}$  at [A4] and a simple expression of I01 is written as:
\begin{equation}
\ I01_{\alpha\beta}(\infty)^{s}=\frac{\kappa_{0}g_{0}\lambda s_{0}\hat{\beta}_{1}}{\hat{\alpha}_{0}(s_{1}+\sigma_{1})+\hat{\beta}_{1}(s_{0}+\sigma_{0})+\mu\hat{\alpha}_{0}\hat{\beta}_{1}}\nonumber\\
\end{equation}

Where the super index \textit{s} it refer to \textit{simple}.
In order to determine $\kappa_{0}$ we use the equality:

\begin{equation}
\int_{0}^{1}\int_{0}^{1}I01_{\alpha\beta}(\infty)^{s}d\hat{\alpha}_{0}d\hat{\beta}_{1}=
\int_{0}^{1}\int_{0}^{1}I01_{\alpha\beta}(\infty)^{c}d\hat{\alpha}_{0}d\hat{\beta}_{1}\nonumber\\
\end{equation}

Substituting  [A9] and [A10] in [A11], considering  $s_{0}$ = $s_{1}$  and $\mu << s_{i}$,

\begin{eqnarray}
\int_{0}^{1}\int_{0}^{1}\frac{\kappa_{0}\ g_{0} \lambda\ d\hat{\alpha}_{0}\ d\hat{\beta}_{1}}{(\lambda+1) (\hat{\alpha}_{0}+ \beta_{1})}=
\int_{0}^{1}\int_{0}^{1}\frac{g_{0}\hat{\beta}_{1}}{(\lambda+1) (\hat{\alpha}_{0}+\beta_{1})}\nonumber\\
\{\frac{2 a_{0} T_{1}}{\lambda}+\kappa_{0} \lambda \{\frac{a_{0}T_{1} \hat{\alpha}_{0} \hat{\beta}_{1}}{\lambda+1} ln(1+\frac{1}{\hat{\beta}_{1}}) ln(\frac{1}{\hat{\alpha}_{0}}+1)\nonumber\\
+\frac{a_{0}T_{1} \hat{\alpha}_{0} \hat{\beta}_{1}}{\lambda+1} ln(\frac{1}{\hat{\beta}_{1}}+1) ln(1+\frac{1}{\hat{\alpha_{0}}})\nonumber\\
\frac{\lambda \hat{\alpha}_{0}}{2 (\lambda+1)}ln(\frac{1}{\hat{\alpha}_{0}}+1)+\nonumber\\
\frac{\lambda \hat{\beta}_{1}}{2 (\lambda+1)} ln(\frac{1}{\hat{\beta}_{1}}+1)\}\} d\hat{\alpha}_{0}\ d\hat{\beta}_{1}\nonumber \\
\end{eqnarray}

Solving the integrals $\kappa_{0}$ is written as:

\begin{eqnarray}
\kappa_{0}=\frac{4 a_{0} (\lambda+1) T_{1}}{\lambda^{2} J_{0}}\nonumber \\
\end{eqnarray}

Substituting  [A13] in [A9] and [A10], the simple and complete expressions for I01 are obtained respectively.


\end{document}